\documentstyle[aps,pra]{revtex}

\begin{document}
\draft

\def\vr{\vec r}
\def\vrp{{\vec r}\,'}
\def\allvr{\{\vr_j\}}
\def\etal{{\it et al. }}
\def\vnabla{{\vec \nabla}}
\def\stop{\nonumber   \\}
\def\rhat{\hat r}
\def\wxc{ W_{xc}}
\def\ex{ {\vec {\cal{ E}}}_{x}}
\def\exr{ {\vec {\cal{ E}}}_{x}(\vr)}
\def\exc{ \vec {\cal {E}}_{xc}}
\def\ee{ \vec {\cal {E}}} 
\def\modrrp{ \mid\vr-\vrp\mid}
\def\half{ \frac{1}{2}}
\def\density{ n(\vr)}
\def\Rnlp{ R_{n' l'}}
\def\Rnl{ R_{n l}}
\def\rlt{r_<}
\def\rgt{r_>}
\def\rltgt{\frac{ \rlt^{l''}} { \rgt^{l''+1} }}
\def\rrp{ \frac{1}{ \mid \vr - \vrp \mid}}

\title{  Positron and positronium affinities in the Work Formalism
Hartree-Fock Approximation}

\author{  Rajendra R. Zope }

\address{  Center of Advanced Studies in Materials Science and Solid State Physics,\\
Department of Physics, University of Pune, Pune-411007, India}

\date{\today}

\maketitle

\begin{abstract}

        Positron binding to anions is investigated within the work formalism 
proposed by Harbola and Sahni (HS) for the halide anions and the systems 
$Li^-$ through $O^-$ excluding $Be^-$ and $N^-$. 
The total ground state energies of the anion-positron  bound systems
are empirically found to be an upper bound to the Hartree-Fock(HF) energies.
The computed expectation values as well as positron and 
positronium affinities are in good  agreement with  their
restricted Hartree-Fock counterparts. 
Binding of a positron to neutral atomic species is also investigated using 
an iterative method.  
\end{abstract}


\section{ Introduction}

     Despite extensive studies [1-11] the problem of a positron forming a 
transient bound state with a neutral atom has not yet been decisively settled. 
It is known that a negatively charged ion always binds a positron;
in fact it turns out that there is an infinite number of bound states.  
The classical Coulomb interaction between an anion and a positron is 
sufficient to bind a positron while in case of neutral atoms the 
polarization of electron charge
distribution is found to be vital for binding \cite{P_C}.
Schrader \etal \cite{Schra_1,Schra_2} have calculated the 
positronium  affinities for halogen atoms using a diffusion quantum Monte
Carlo technique in which the core region is represented by a model
potential.  Accurate calculations on the  positronium 
hydride, positronium helium and positronium lithium systems using the all
particle diffusion Monte Carlo  method have been recently reported 
by Yoshida \etal\cite{Li_1,Li_2}. Very recently, Bressanini\cite{Li_4}
\etal have investigated the positron-anion bound states
for the anionic species  $Li^-,B^-,C^-,O^-$ and $F^-$ using variational
and diffusion quantum Monte Carlo techniques. They found that except $B^-$
all these systems are stable against dissociation into the 
corresponding neutral atom and positronium.
Such methods  are computationally very demanding thereby restricting
 their applicability to systems with fewer particles. 
The independent  particle approximation methods  such as the Hartree-Fock 
or density functional theories enable one to compute with ease the anionic-positron
interactions for all the anions in the  periodic table.
  Calculations within the restricted HF theory 
have been reported by Cade and Farazdel\cite{Cade1,Cade3}
and Patrick and Cade\cite{P_C} for the systems $Li^-$ through $O^-$ as well as 
for halogen anions. They have also examined the stability of anion-positron bound
systems with respect to their dissociation into the corresponding  atom and  positronium.
For the description of many electron-positron bound system within the density functional
theory (DFT), Chakraborty\cite{chakra} has proposed a two component density  functional 
theory wherein the electron and the positron densities play the role of
basic variables.
This was subsequently extended by Harrison \cite{harrison} to incorporate 
self-interaction
correction\cite{SIC} wherein the calculated positron and positronium affinities are 
in close agreement with their HF counterparts. 
By invoking the Slater transition state concept Baruah \etal \cite{Tunna_1}
have obtained decent estimates of positron affinities to negative ions.  
Earlier, density functional calculations of  the positron-anion bound
state had also been carried out  by Kanhere \etal \cite{DGK}.

Recently,  an attractive alternative to the Hartree-Fock theory was proposed
by Harbola and Sahni \cite{HS_prl}.  The exchange-correlation potential 
in the so-termed work formalism of Harbola-Sahni (HS) is  obtained as the 
work done in bringing an electron in the electric field of its 
Fermi-Coulomb hole charge distribution.
The work formalism HF approximation essentially emerges if one ignores
Coulomb correlations and the correlation contribution to the kinetic
energy in the Kohn-Sham theory \cite{Sahni}. Recently, Holas and March
\cite{H_M} have shown that the HS potential can be obtained from the 
full second order density  matrix.
The total atomic energies and  various one electron properties 
of work  formalism are practically equivalent to those of HF 
theory\cite{HS_92}.
Unlike other local exchange-only density functionals, the work  formalism
of HS selectively  gives  convergent orbitals and eigenvalues for negative
 ions that are comparable
to the HF accuracy\cite{sen}. In the present communication, we address 
the problem of positron binding to anions and atoms 
within the exchange-only work formalism. 
The purpose of the present work is two-fold: $1)$ To compute the positron
affinities and binding energies using {\em local orbital-independent} density
functional theory.  $2)$ To test the work formalism of the
Harbola-Sahni  for the description of many electron-positron system.
The computed positron affinities, 
binding energies and $<\!r^n\!>-$ moments ($n=-1$ through $2$) within the work formalism
are compared against their restricted Hartree-Fock (RHF) counterparts.
For the neutral atom-positron bound states we  use an iterative method 
similar to the one used by Patrick and Cade\cite{P_C}.  The  
plan on the presentation is as follows: In section II we outline the theory
of work formalism for the description of electron-positron system while 
section III deals with the results and discussion which will be followed by 
conclusions in section IV.

\section{Theory}

    The non-relativistic Hamiltonian $H$ for an N-electron and one positron 
system is  the sum of the electronic part $H_e$ (atomic units are used throughout),
\begin{equation}
H_e  =  \sum_i^N \Biggr ( -\frac{\nabla^2_i}{2} \Biggr ) 
- \sum_i^N \frac{Z}{r_i}
      +  \sum_{i<j}^N \frac{1}{\mid\!\vr_i-\vr_j\!\mid}
\end{equation}
consisting of  the kinetic energy
of electrons, the electron-nuclear interaction, the electron-electron
repulsion,
and   the Hamiltonian $H_p$ for a positron,
\begin{equation}
H_p  =   -\frac{\nabla^2_p}{2} +  \frac{Z}{r_p}  
      -  \sum_{i=1}^N \frac{1}{\mid\!\vr_i-\vr_p\!\mid}
\end{equation}
containing  the positron 
kinetic energy, the nucleus-positron repulsion and the positron-electron
attractive interaction.
%

Within the exchange-only work formalism of Harbola-Sahni the 
local exchange potential  is obtained as the work done in moving an
electron in the electric field $\cal{\vec{E}}$  of its Fermi hole 
$\rho_x(\vr,\vrp)$ charge distribution, that is,
\begin{equation}
W_x(\vr) = - \int_{\infty}^{\vr}  \ex(\vrp)\cdot d\vec{l}~'
\end{equation}
with 
\begin{equation}
\exr = \int \frac{\rho_x(\vr,\vrp)}{\modrrp^3}(\vr-\vrp)
d^3r'
          \label{ex:eq}
\end{equation}
and  
\begin{equation}
\rho_{x}(\vr,\vrp) =  -\frac{\sum_i \phi_i^*(\vr)\phi_i(\vrp)\phi_j^*(\vrp)\phi_j(\vr)}
                      {\sum_k\phi_k^*(\vr) \phi_k(\vr)}.
          \label{rhox:eq}
\end{equation}
Now, in order that the effective potential experienced by electron is well
defined the curl of the ``exchange electric field''  represented by Eq.
(\ref{ex:eq})  should vanish. This is the case for the closed shell
atoms and open shell atoms in the central field approximation in
which the present calculations have been carried out (see references  
\cite{Sahni}, \cite{HS_92}for details). It is to be the noted that HS exchange 
potential  obtained this way differ from the exact Kohn-Sham potential
only by the kinetic correlation contribution \cite{L_M,Sahni}.
The orbitals, $\phi_i$, in Eq. (\ref{rhox:eq}) are the solutions of
the (Kohn-Sham like) HS equation
\begin{eqnarray}
 \Biggl \{ -\frac{\nabla^2}{2}  - \frac{Z}{r} + \int \frac{n^-(\vrp)
}{\mid\!\vr-\vrp\!\mid} d^3r'+ W_x(\vr) 
 - \int \frac{n^+(\vrp) }{\mid\!\vr-\vrp\!\mid} d^3r' 
\Biggr \}
\psi_i(\vr)
= \epsilon_i   
\psi_i(\vr).
           \label{scf1:eq}
\end{eqnarray}
Here, $n^-(\vr)=\sum_i^N \mid\!\psi_i(\vr)\!\mid^2$ is the electron
density, $n^+(\vr)= \mid\!\psi^+(\vr)\!\mid^2$ is the positron density with
$\psi^+$ being the solution of the corresponding 
differential equation for the positron 
\begin{eqnarray}
 \Biggl \{ -\frac{\nabla^2}{2}  + \frac{Z}{r} - \int \frac{n^-(\vrp)
}{\mid\!\vr-\vrp\!\mid} d^3r'   \Biggr \}  \psi^+(\vr)
  = \epsilon^+ \psi^+(\vr).
           \label{scf2:eq}
\end{eqnarray}
The effective potential seen by the  positron in the exchange-only 
formalism also has the interpretation  as
the work done in moving the positron in the field  
of the electronic and nuclear charge distribution. 
Eqs. (\ref{scf1:eq}) and (\ref{scf2:eq}) are  solved self consistently 
to obtain the ground state 
energy of the electron-positron combined system which is expressed as
\begin{eqnarray}
E = T_e + T_p 
- Z \int \frac{ n^-(\vr)}{r} ~d^3r 
+ Z \int \frac{ n^+(\vr)}{r}  ~d^3r 
+ E_x
- \int \frac{n^-(\vr) n^+(\vrp) }{\modrrp}d^3r  d^3r'
+\frac{1}{2} \int \frac{n^-(\vr) n^-(\vrp) }{\modrrp} d^3r d^3 r'.
\end{eqnarray}
The first two terms, $T_e$ and $T_p$ denote respectively, the kinetic
energy of electrons and the positron, the next two terms represent the
attractive and the repulsive interaction energies  of the electrons and
a positron with the nuclear charge, $E_x$ is the exchange energy while the
last two terms signify the  electron-electron and electron-positron 
interaction energies respectively. The expressions for the calculation of
exchange energy and electron-electron interaction energies can be found
in the appendix of Ref. \cite{HZP}. The electron-positron energy can be
obtained by following the steps exactly similar to the electron-electron
case and emerges as

 \begin{eqnarray}  
E & = & - \sum_{nlm} \sum_{n'l'm'} N_{n l m} 
\int r^2 ~dr ~r'^2 ~dr'  
~\Rnl^2(r) ~{\Rnlp^+}\!^2(r') \rltgt     \nonumber   \\ 
 &  &  ~~~\times ~~(2l +1)~ (2l'+1) ~\left(\matrix{   l  & l''  &  l
\cr 
                m   &  0   &  -m \cr} \right)  
\left(\matrix{ l   &  l'' &  l \cr 
                0   &   0  &  0  \cr} \right)  \nonumber   \\ 
 &   &  ~~~~\times ~~~\left(\matrix{ l'  & l''  &  l' \cr 
                m'  & 0   &  -m'  \cr} \right)  
  \left(\matrix{ l'  & l''  &  l' \cr 
                0    & 0   &  0  \cr} \right).
                             \label{ep:en}
~~~~~~~~~~~~~~~~~~~~~~~~~~~~~~~~~\end{eqnarray}
Here, $R_{nl}$ and $\Rnlp^+$ are respectively the radial parts of the electron
orbitals and the  positron orbitals, and $N_{nlm}$ is the orbital occupancy. The
$3j$ symbols in Eq. (\ref{ep:en}) arise due  to integration over 
the solid angle $\Omega\equiv \Omega(\theta,\phi)$.

The Herman-Skillman code\cite{HS:1}, modified for the Harbola-Sahni
potential has been  further modified in order to incorporate the positron.
The calculations are carried out in the {\em central}-field
approximation for the systems $Li^-$ through $F^-$ 
except for $Be^-$ and $N^-$ and the halide ions.  In order to obtain the
ground state of an anion $A^-$, we start with 
converged potential of the neutral atom $A$ and perform self-consistent 
calculation.  The converged potential of
the anionic system  $A^-$ was then taken as a starting potential for the
anion-positron self-consistent calculation.
 This was done  in order to achieve 
fast convergence.  In the following section we present our main
results.

\section{Results}

The total energies of the anion-positron bound states  calculated
in the present formalism and the corresponding restricted HF (RHF)
energies for the positron in different states are displayed 
in Table \ref{TE:tab}. The RHF  numbers
for total energies, positron and positronium affinities 
against which we compare our results are due to Patrick and Cade\cite{P_C}
and Cade and Farazdel\cite{Cade1}.
The present total energies are in good agreement  with the HF energies.
The differences in parts per million between the energies of the present 
work and those of the HF theory are given in Table \ref{ppm:tab} for 
the anion-positron bound state (the positron is in the $1s$ orbital).
The differences  diminish with the size of the anion.
It is also evident from the table that the calculated total energies are 
slightly higher than the HF energies.  This is expected since the HS orbitals
differ from the HF orbitals which variationally minimize the total energy.

The positron affinity  is defined as 
\begin{equation}
P.A. = E(A^-) + E(e^+) - E([A^-;e^+]),
                   \label{pa:eq}
\end{equation}
where  $E([A^-;e^+])$ denotes the energy of the anion-positron bound system.
Positive value of the P.A. indicates that $E[A^-;e^+]$ is a bound state, 
that is, 
the system $A^-$ will bind a positron.  The calculated positron affinities
and the negative of the positron eigenvalue are given Table in \ref{PA:tab},
also given are the corresponding HF values given for comparison.
For all the systems, the PA is positive indicating their stability with
respect to dissociation into an anion and a positron.  
The present values of the positron energy eigenvalues are in general 
higher in magnitude than the corresponding HF values and  lie
between the PA calculated
from Eq.  (\ref{pa:eq}) and the recent diffusion quantum Monte Carlo
(QMC) values \cite{Li_4}. The available diffusion Monte Carlo
 values\cite{Li_4} for the 
positron affinities for different systems (in ground state) are   $6.507 (Li^-)$,
$6.015 (B^-)$,  $5.941 (C^-)$, $5.862 (O^-)$  and $6.170 (F^-)$ in $eV$.    

Another binding energy of importance is the {\em positronium} affinity, a
positive value of which means the system $[A^-;e^+]$ is stable
with respect to break up into the positronium $(Ps)$ and a neutral atom
$A$.
The binding energies or positronium affinities can be computed in various
ways \cite{Schra_r}. We compute the positronium affinities using the following
two definitions:
\begin{eqnarray}
  && PsA = E(A) + E(Ps) - E([A^-;e^+]) 
                 \label{ps_1:eq}   \\
\mbox{and}&&  \nonumber \\
  && PsA = E.A. + P.A. + E(Ps) .
                 \label{ps2:eq}
\end{eqnarray}
Positronium affinities computed from Eq.(\ref{ps_1:eq}) are compared
against   the restricted Hartree-Fock (RHF)  positronium affinities
\cite{P_C,Cade1}  in Table \ref{PA:tab}. 
While for all the systems investigated the present and the RHF values of
positronium affinities (calculated using Eq. (\ref{ps_1:eq}))
are in good agreement,  no system is stable
with respect to dissociation into a neutral atom and positronium:
the positronium affinities for all systems  are negative.

In order to calculate the positronium affinity using Eq. (\ref{ps2:eq}),
we choose $-\epsilon_{max}$ of anionic system for the electron affinity (EA) 
as it is empirically found \cite{kli} that in the 
exchange-only work formalism and the HF theory, the $-\epsilon_{max}$ of 
the anionic system is, in general, a better estimate of EA than those
obtained from the difference of self-consistent total energies of the
atom and the corresponding anion. Further, it is observed that 
such estimates of EA within the present formalism are
closer to the experimental EA \cite{hotop} than those obtained in HF theory
by means of  Koopmans' theorem\cite{sen}.  
For positron affinity 
we employ $-\epsilon^+$ (third column in Table \ref{PA:tab}) 
since this quantity is, in general, in better agreement with 
the accurate QMC positron affinity than the 
one obtained by taking  difference of self-consistent energies
(using Eq. (\ref{pa:eq})).
The positronium affinities thus calculated are also  given in  the last column of the Table
 \ref{PA:tab}. 
  These values of $PsA$ are less negative than the $PsA$ computed
as the difference of the self-consistent energies (Eq.(\ref{ps_1:eq})
leading in some cases to the binding.
The systems $[O^-;e^+_{1s}]$, $[F^-;e^+_{1s}]$, $[F^-;e^+_{2p}]$,
$[Cl^-;e^+_{1s}]$ and  $[Br^-;e^+_{1s}]$ are found to be stable 
against the dissociation into the positronium and an atom. This  binding
may be attributed to the accurate asymptotic structure of the work formalism
HF approximation.

  We finally present the one electron properties such as $<\!r^n\!>$
expectation values for halide anions
in the Table \ref{exp:tab}.   The one electron expectation values are in good
agreement with their HF counterparts.
The computed $\!<1/r\!>$ values are slightly 
 larger than the HF values, implying the slight increase 
in the positron density towards the nucleus which therefore,
should be compensated by small reduction in the long-range
of the positron density leading to smaller $\!<r\!>$ and $\!<r^2\!>$.
This is indeed the case as can be seen from  the Table \ref{exp:tab}.

We have also investigated the binding of positron to neutral atoms
 in the spirit of
Patrick and Cade, by starting  with  the anion-positron bound state and 
reducing the ionicity of the system to obtain desired neutral system.
It was  found that the neutral-atom positron bound state 
does not exist in the exchange-only work formalism. This, 
however, is   not surprising as the present treatment 
lacks the electron-positron and electron-electron correlation 
effects which are crucial to permit such a binding\cite{Tunna_1}.
Further, it has been rigorously shown by Pathak\cite{pathak} 
that the deviation from the spherical symmetry is a necessary condition 
 in order that the positron binding to neutral atom would occur.
The present treatment can be extended to include  the correlations effects,
namely the electron-electron Coulomb correlations, correlation
contribution to the kinetic energy and the  electron-positron correlation.
The first one can be incorporated by modeling the correlation second-order density 
matrix as suggested by Levy and Perdew \cite{L_M} while the second one can be
derived in terms of density matrices via virial theorem \cite{H_M}.
The electron-positron correlation potential\cite{Puska} can be added in 
an {\em ad hoc} manner to the effective potential.

\section{Conclusions}
  In the present work, the positron binding to negative ions is
investigated within the exchange-only work formalism. 
The work formalism of Harbola-Sahni seems to  provide the Hartree-Fock level
description of the electron-positron system as can be seen from the agreement
between the present values of positron and positronium affinities and 
their restricted Hartree-Fock counterpart.  The advantage of the 
work formalism is that its effective potential is local,  orbital
independent and therefore computationally cheaper. 
The systems $[O^-;e^+_{1s}]$, $[F^-;e^+_{1s}]$, $[F^-;e^+_{2p}]$,
$[Cl^-;e^+_{1s}]$ and  $[Br^-;e^+_{1s}]$ are found to be stable 
against the dissociation into the positronium and the corresponding atom.

\newpage

\begin{table}

\caption{
    Negative total energies (in Hartree a.u.) of $[A^-;e^+]$ bound system calculated
 in central field approximation within the work formalism. Also given are
 the total energies in HF theory.}

\begin{tabular}{lcclccrl}
{System} & Work formalism   &     HF         & \hskip 0.7in {System}  & Work formalism   &     HF  & & \\
         &   $-E$           &  $-E$          & \hskip 0.7in           &  $-E$            &  $-E$   & &
\\ \hline \\
$Li^-$ $1s^2 ~2s^2$ & 7.4270 &   7.4282      & \hskip 0.7in $F^-$ $1s^2 2s^2 2p^6$ &  99.4543  &  99.4594 & \\
$Li^-$  $:1s$    & 7.5286     &   7.5299     & \hskip 0.7in $F^-$ $:1s$     &  99.6383    &     99.6434   & \\
$Li^-$  $:2s$    & 7.4748     &   7.4760     & \hskip 0.7in $F^-$ $:2s$     &  99.5253    &     99.5305   & \\
$Li^-$  $:3s$    & 7.4528     &   -          & \hskip 0.7in $F^-$ $:2p$     &  99.5641    &     99.5692   & \\
$Li^-$  $:2p$    & 7.5017     &   7.5030     & \hskip 0.7in $F^-$ $:3s$     &  99.4917    &      -   & \\
$Li^-$  $:3p$    & 7.4653     &   -          & \hskip 0.7in $F^-$ $:3p$     &  99.5048    &      -   & \\
$Li^-$  $:3d$    & 7.4752     &   7.4765     & \hskip 0.7in $F^-$ $:3d$     &  99.5095    &     99.5147   & \\
$Li$  $1s^2 ~2s^1$  &  7.4316    &    7.4328    & \hskip 0.7in $F$   $1s^2 2s^2 2p^5$   &  99.4046    &     99.4095   & \\
$B^-$ $1s^2 2s^2 2p^2$  &   24.5156    &     24.5192   & \hskip 0.7in $Cl^-$ $[Ne]3s^2 3p^6$  &   459.5640   &    459.5769   & \\
$B^-$ $:1s$    &   24.6495    &     24.6531  & \hskip 0.7in $Cl^-$ $:1s$   &   459.7071   &    459.7189   & \\
$B^-$ $:2s$    &   24.5733    &     24.5769  & \hskip 0.7in $Cl^-$ $:2s$   &   459.6243   &    459.6373   & \\
$B^-$ $:2p$    &   24.6108    &     24.6202  & \hskip 0.7in $Cl^-$ $:2p$   &   459.6625   &    459.6754   & \\
$B^-$ $:3s$    &   24.5477    &     -        & \hskip 0.7in $Cl^-$ $:3s$   &   459.5972   &    -   & \\
$B^-$ $:3p$    &   24.5610    &     -        & \hskip 0.7in $Cl^-$ $:3p$   &   459.6107   &    -   & \\
$B^-$ $:3d$    &   24.5694    &     24.5757  & \hskip 0.7in $Cl$   $[Ne]3s^2 3p^5$  & 459.4697     &    459.4830   & \\
$B$   $1s^2 2s^2 2p^1$  &   24.5261    &     24.5292   & \hskip 0.7in $Br^-$  $[Ar]4s^23d^{10} 4p^6$ &  2572.523 &    2572.5363   & \\
$C^-$ $1s^2 2s^2 2p^3$  &   37.7041    &     37.7088   & \hskip 0.7in $Br^-$ $:1s$     & 2572.656   &     2572.6695   & \\
$C^-$ $:1s$    &   37.8563    &     37.8610  & \hskip 0.7in $Br^-$ $:2s$     & 2572.5803  &     -   & \\
$C^-$ $:2s$    &   37.7671    &     37.7718  & \hskip 0.7in $Br^-$ $:2p$     & 2572.6177  &     2572.6311   & \\
$C^-$ $:2p$    &   37.8040    &     37.8087  & \hskip 0.7in $Br$   $[Ar]4s^23d^{10} 4p^5$  & 2572.229 &     -   & \\
$C^-$ $:3s$    &   37.7384    &     -        & \hskip 0.7in & &  & \\
$C^-$ $:3p$    &   37.7513    &     -        & \hskip 0.7in & &  & \\
$C^-$ $:3d$    &   37.7584    &     37.7632  & \hskip 0.7in & &  & \\
$C$   $1s^2 2s^2 2p^2$  &   37.6847    &     37.6887   & \hskip 0.7in & &  & \\
$O^-$ $1s^2 2s^2 2p^5$   &  74.7849    &     74.7897   & \hskip 0.7in & &  & \\
$O^-$ $:1s$     &  74.9583    &     74.9630  & \hskip 0.7in & &  & \\
$O^-$ $:2s$     &  74.8534    &     74.8582  & \hskip 0.7in & &  & \\
$O^-$ $:2p$     &  74.8940    &     74.9026  & \hskip 0.7in & &  & \\
$O^-$ $:3s$     &  74.8214    &     -        & \hskip 0.7in & &  & \\
$O^-$ $:3p$     &  74.8350    &     -        & \hskip 0.7in & &  & \\
$O^-$ $:3d$     &  74.8403    &     74.8461  & \hskip 0.7in & &  & \\
$O$   $1s^2 2s^2 2p^4$   &  74.8050    &     74.8095 & \hskip 0.7in & &    & \\
\end{tabular}
\label{TE:tab}
\end{table}

\begin{table}

\caption{
    The total ground state energy differences between the present
work and the HF theory in parts per million for anion-positron bound states
(the positron is in the $1s$ orbital).}

\begin{tabular}{cccc}
  system   &       differences(ppm) \\
\hline
  $[Li^-,e^+_{1s}]$   &    173      \\
  $[B^-,e^+_{1s}]$    &    146     \\
  $[C^-,e^+_{1s}]$    &    124    \\
  $[O^-,e^+_{1s}]$    &    59     \\
  $[F^-,e^+_{1s}]$    &    51      \\
  $[Cl^-,e^+_{1s}]$   &    26       \\
  $[Br^-,e^+_{1s}]$   &    5       \\
\end{tabular}
      \label{ppm:tab}
\end{table}

\newpage

\begin{table}

\caption{
    Positron eigenvalues,affinities and positronium affinities in $eV$ }      
\begin{tabular}{ccccccccc}
system & $n_+l_+$ & \multicolumn{2}{c}{$-\epsilon^+$} & \multicolumn{2}{c} {PA} &  \multicolumn{2}{c}{PsA} & PsA$^{\dagger}$ \\
       &          &   HS &  RHF                       &   HS &  RHF             &     HS &        RHF      &   \\
\hline
$Li^-$ & $:1s$ &  3.049 &  2.996 &  2.765 &  2.766 & -4.161 & -4.159&-3.341 \\
$Li^-$ & $:2s$ &  1.347 &  1.329 &  1.306 &  1.301 & -5.625 & -5.624&-5.0427 \\
$Li^-$ & $:3s$ &  0.770 &    -   &  0.770 &    -   & -6.169 &    -  &-5.620 \\
$Li^-$ & $:3p$ &  1.064 &    -   &  1.042 &    -   & -5.883 &    -  &-5.326  \\
$Li^-$ & $:3d$ &  1.339 &  1.331 &  1.312 &   1.312& -5.614 & -5.613&-5.051  \\
$B^-$ & $:1s$  & 3.785 &  3.778  & 3.644   & 3.642 & -3.422 & -3.428&-2.395 \\
$B^-$ & $:2s$  & 1.584 &  1.582  & 1.714   & 1.569 & -5.372 & -5.502&-4.596  \\
$B^-$ & $:2p$  & 2.557 &  2.799  & 2.590   & 2.748 & -4.240 & -4.323&-3.623  \\
$B^-$ & $:3s$  & 0.876 &    -    & 0.875   &   -   &    -   &    -   &-5.304 \\
$B^-$ & $:3p$  & 1.212 &    -    & 1.235   &   -   &    -   &    -   &-4.968 \\
$B^-$ & $:3d$  & 1.442 &   1.540 & 1.464   &  1.536&  -5.450&  -5.535&-4.738 \\
$C^-$ & $:1s$ &  4.204 &   4.218&  4.142   &  4.141&  -2.131&  -2.21& -0.854 \\
$C^-$ & $:2s$ & 1.712  &  1.718 & 1.714    & 1.713 & -4.558 & -4.540& -3.346 \\
$C^-$ & $:2p$ & 2.712  &  2.135 & 2.718    & 2.718 & -3.554 & -3.535& -2.346 \\
$C^-$ & $:3s$ & 0.931  &   -    & 0.933    &  -    & -5.339 &   -   & -4.127 \\
$C^-$ & $:3p$ & 1.282  &   -    & 1.284    &  -    & -4.988 &   -   & -3.776 \\
$C^-$ & $:3d$ & 1.478  &  1.480 & 1.478    & 1.480 & -4.795 & -4.773&  -3.580 \\
$O^-$ & $:1s$ & 4.784  & 4.769  & 4.718    & 4.716 & -2.628 & -2.662&  1.443  \\
$O^-$ & $:2s$ & 1.869  & 1.865  & 1.864    & 1.862 & -5.483 & -5.475& -1.472  \\
$O^-$ & $:2p$ & 2.917  & 3.079  & 2.969    & 3.070 & -4.378 & -4.268& -0.424  \\
$O^-$ & $:3s$ & 0.993  &   -    & 0.993    &  -    & -6.354 &   -   & -2.348 \\
$O^-$ & $:3p$ & 1.350  &   -    & 1.363    &  -    & -5.984 &   -   & -1.991 \\
$O^-$ & $:3d$ & 1.498  & 1.532  & 1.508    & 1.532 & -5.839 & -5.805& -1.843  \\
$F^-$ & $:1s$ & 5.061  & 5.048  & 5.007    & 5.006 & -0.441 & -0.434  &3.126  \\
$F^-$ & $:2s$ & 1.932  & 1.936  & 1.932    & 1.933 & -3.516 & -3.506  &0.003  \\
$F^-$ & $:2p$ & 2.993  & 2.992  & 2.998    & 2.987 & -2.460 & -2.452  &1.058  \\
$F^-$ & $:3s$ & 1.020  &   -    & 1.018    &  -    & -4.430 &   -     &-0.015 \\
$F^-$ & $:3p$ & 1.377  &   -    & 1.374    &  -    & -4.073 &   -     &-0.598 \\
$F^-$ & $:3d$ & 1.503  & 1.502  & 1.503    & 1.503 & -3.946 & -3.937  &-0.432  \\
$Cl^-$ & $:1s$ & 3.928 &  3.922 &  3.948   &  3.894 & -0.340&  -0.350 & 1.052  \\
$Cl^-$ & $:2s$ & 1.645 &  1.644 &  1.641   &  1.642 & -2.590&  -2.600 &-1.231  \\
$Cl^-$ & $:2p$ & 2.687 &  2.687 &  2.680   &  2.680 & -1.554&  -1.560 &-0.189  \\
$Br^-$ & $:1s$ & 3.655 &  3.653 &  3.619   &  3.626 & -0.620&  -0.690 & 0.409  \\
$Br^-$ & $:2s$ & 1.568 &    -   &  1.559   &    -   & -2.680&    -    &-1.678  \\
$Br^-$ & $:2p$ & 2.585 &  2.588 &  2.577   &  2.577 & -1.670&  -1.740 &-0.661  \\ 
\end{tabular}
      \label{PA:tab}
\end{table}
$^\dagger$ Calculated using Eq. (\ref{ps2:eq}) (see text for details).

\newpage

\begin{table}
\caption{
The radial expectation values $<\!\frac{1}{r}\!>$,
 $<\!{r}\!>$ and $<\!{r^2}\!>$ for the $1s$ positron orbital in halide ions
calculated within the work formalism and  the corresponding Hartree-Fock
values (the values are in Hartree atomic units).} 

\begin{tabular}{ccccccc}
%
system  &  \multicolumn{2}{c} { $<\!\frac{1}{r}\!>$ }  &  \multicolumn{2}{c}{  $<\!{r}\!>$ }    &  \multicolumn{2}{c}    {   $<\!{r^2}\!>$ } \\  
        &       HS   &     HF   &    HS  &     HF  &       HS   &   HF  \\
\hline
$[F^-;e^+]$ &0.2959  &  0.2948  & 4.067  &  4.080  &    19.454  & 19.572 \\
$[Cl^-;e^+]$  & 0.2194  &  0.2189  & 5.2987 &  5.3085 & 32.2031    & 32.3131  \\
$[Br^-;e^+]$ &0.2018 &  0.2015  & 5.7178 &  5.7198 &    37.1800 & 37.2715 \\
\end{tabular}
        \label{exp:tab}
\end{table}

\begin{acknowledgments}
The author gratefully acknowledges  the help of Dr. M. K.  Harbola
for providing him with a Herman-Skillman work formalism code. 
The author would also like to thank Dr. R K. Pathak, Dr. A. Kshirsagar
and Ms. Tunna Baruah for the helpful discussions and 
encouragement, and  further  acknowledge the Council of Scientific and
Industrial Research, New Delhi, India for the assistance for this
work in the form of senior research fellowship.
\end{acknowledgments}

\end{document}